\documentclass[twocolumn,aps,prd]{revtex4}
 \usepackage{graphicx,amssymb}
 \usepackage[utf8]{inputenc} 
 
\usepackage[urlcolor=blue,colorlinks,breaklinks=true]{hyperref}

\PassOptionsToPackage{obeyspaces,spaces,hyphens}{url}
 
 \usepackage{amsmath}

\begin{document}
 	\def\half{{1\over2}}
 	\def\shalf{\textstyle{{1\over2}}}
 	
 	\newcommand\lsim{\mathrel{\rlap{\lower4pt\hbox{\hskip1pt$\sim$}}
 			\raise1pt\hbox{$<$}}}
 	\newcommand\gsim{\mathrel{\rlap{\lower4pt\hbox{\hskip1pt$\sim$}}
 			\raise1pt\hbox{$>$}}}

\newcommand{\be}{\begin{equation}}
\newcommand{\ee}{\end{equation}}
\newcommand{\bq}{\begin{eqnarray}}
\newcommand{\eq}{\end{eqnarray}}

\title{Halos of dark energy}
 	 	
\author{P.P. Avelino}
\affiliation{Departamento de F\'{\i}sica e Astronomia, Faculdade de Ci\^encias, Universidade do Porto, Rua do Campo Alegre 687, PT4169-007 Porto, Portugal}
\email[Electronic address: ]{pedro.avelino@astro.up.pt}
\affiliation{Instituto de Astrof\'{\i}sica e Ci\^encias do Espa{\c c}o, Universidade do Porto, CAUP, Rua das Estrelas, PT4150-762 Porto, Portugal}
\affiliation{Centro de Astrof\'{\i}sica da Universidade do Porto, Rua das Estrelas, PT4150-762 Porto, Portugal}

\date{\today}
\begin{abstract}
We investigate the properties of dark energy halos in models with a nonminimal coupling in the dark sector. We show, using a quasistatic approximation, that a coupling of the mass of dark matter particles to a standard quintessence scalar field $\phi$ generally leads to the formation of dark energy concentrations in and around compact dark matter objects. These are associated with regions where scalar field gradients are large and the dark energy equation of state parameter is close to $-1/3$. We find that the energy and radius of a dark energy halo are approximately given by $E_{\rm halo} \sim \boldsymbol{\beta}^2 \varphi \,  m$ and $r_{\rm halo} \sim  \sqrt{\boldsymbol{\beta} \,\varphi ({R}/{H})}$, where $\varphi=Gm/(R c^2)$, $m$ and $R$ are, respectively, the mass and radius of the associated dark matter object, $\boldsymbol{\beta} = -(8\pi G)^{-1/2} d \ln m/d \phi$ is the nonminimal coupling strength parameter, $H$ is the Hubble parameter, $G$ is the gravitational constant, and $c$ is the speed of light in vacuum. We further show that current observational limits on $\boldsymbol{\beta}$ over a wide redshift range lead to stringent constraints on $E_{\rm halo}/m$ and, therefore, on the impact of dark energy halos on the value of the dark energy equation of state parameter. We also briefly comment on potential backreaction effects that may be associated with the breakdown of the quasistatic approximation and determine the regions of parameter space where such a breakdown might be expected to occur.
\end{abstract} 

\maketitle
 	
\section{Introduction}
\label{sec:intr}

In general relativity an exotic dark energy (DE) fluid \cite{Frieman:2008sn,Caldwell:2009ix,Li:2011sd} dominating the energy density of the Universe is required in order to explain the recent acceleration of the expansion of the Universe  \cite{BOSS:2016wmc,Pan-STARRS1:2017jku,Planck:2018vyg}. Various observations also suggest that  matter in the Universe is primarily nonbaryonic and dark \cite{Cooke:2017cwo,Planck:2018vyg}. Despite their importance for the dynamics of the Universe, the physical nature of nonbaryonic dark matter (DM) and DE remains largely unknown. In particular, it is not known whether or not DM and DE are nonminimally coupled \cite{Wetterich:1994bg,Amendola:1999er,Zimdahl:2001ar,Farrar:2003uw}, or even if they could be associated with a single DE fluid \cite{Avelino:2003cf,Avelino:2008zz}. It is also possible that general relativity may not provide an accurate description of gravity on cosmological scales, and that DE and/or DM could be manifestations of modified gravity \cite{Nojiri:2010wj,Clifton:2011jh,Avelino:2016lpj}.

The coupling of the mass of DM particles to a DE scalar field \cite{Amendola:1999er,Zimdahl:2001ar,Farrar:2003uw} has been shown to give rise to DE mediated fifth forces between DM particles as well as velocity dependent forces. These forces can affect the linear growth of cosmological perturbations \cite{Pettorino:2008ez,CalderaCabral:2009ja,Baldi:2010pq}, and may also play a crucial role on nonlinear scales with a potential impact on the dynamics of galaxies and clusters of galaxies \cite{Manera:2005ct,Mainini:2006zj,Abdalla:2007rd,Abdalla:2009mt,Avelino:2011zx,Baldi:2011wy,Cui:2012is}. Also, a nonminimal coupling between DM and DE has been claimed to alleviate some cosmic tensions \cite{Planck:2015bue,DiValentino:2017iww,Barros:2018efl,Yang:2018euj,Pan:2019gop,DiValentino:2019ffd}, including the apparent discrepancy between local and high redshift constraints on the value of the Hubble parameter, making this a promising avenue of research.

Although many studies do not explicitly consider them, local variations in the DE have been shown to play an important role in the context of growing neutrino models, with a potentially significant backreaction on the background evolution of the Universe \cite{Ayaita:2011ay,Ayaita:2012xm,Casas:2016duf}. It has also been shown that the dynamics of DM particles nonminimally coupled to a DE field may be affected by backreaction effects with a possible impact on structure formation and on the large scale dynamics of the Universe \cite{Avelino:2015fka}.

This paper aims to characterize the properties of DE halos resulting from a nonminimal coupling between DM and a DE scalar field.  We shall work in the context of the quasistatic approximation, which essentially consists in neglecting terms involving time derivatives in the perturbed field equations. This approximation has been previously used to investigate local variations of the fine-structure constant inside virialized objects in the context of a DE model with a nonminimal coupling to the electromagnetic field \cite{Avelino:2005pw}. It has also been frequently used in the context of other DE and modified gravity scenarios  \cite{Noller:2013wca,Bose:2014zba,Sawicki:2015zya,Pace:2020qpj}. The quasistatic approximation is particularly useful when considering small subhorizon scales, since the dynamics of perturbation modes with a wave number significantly smaller than the Hubble radius is in general expected to be dominated by the terms containing spatial derivatives in the equations of motion. Under these conditions, the quasistatic approximation may significantly simplify the perturbation equations, and allow for an  analytical treatment. 

The outline of this paper is as follows. In Sec. \ref{sec2} we describe a generic family of quintessence models with a nonminimal coupling to the DM and derive the corresponding equations of motion. In Sec. \ref{sec3} we use the quasistatic approximation to compute the scalar field perturbations around compact DM objects, discussing the conditions required for its applicability. In Sec. \ref{sec4} we estimate the energy and radius of a DE halo surrounding a compact DM object, as well as the corresponding DE equation of state parameter. We also provide a lower bound to the contribution of DE perturbations inside compact DM objects. Finally, we discuss the implications of our results and conclude in Sec. \ref{sec:conc}. 

Throughout this paper we use units where the speed of light in vacuum is equal $c=1$.  We also adopt the metric signature $(-,+,+,+)$. The Einstein summation convention will be used when a greek index appears twice in a single term, once in an upper (superscript) and once in a lower (subscript) position.

\section{Nonminimally interacting dark sector \label{sec2}}

In this paper we consider a class of models for the dark sector with a nonminimal coupling between DM and a standard quintessence scalar field (which plays a DE role). These models are described by the action
\be\label{eq:L}
S=\int d^2x \, \sqrt{-g}  \, {\mathcal L}\, ,
\ee
where the Lagrangian $\mathcal L$ is given by
\be
{\mathcal L}=  {\mathcal L}_{\rm DE}+ {\mathcal L}_{\rm DM}\,,
\ee
with
\bq
{\mathcal L}_{\rm DE} &=&  X - V(\phi) \,,\\
{\mathcal L}_{\rm DM} &=&  f(\phi){\mathcal L}_{\rm DM*} \,.
\eq
Here $X=-\nabla^\mu \phi \nabla_\mu \phi/2$ is a standard kinetic term, $V(\phi)$ is the scalar field potential, 
\be
{\mathcal L}_{\rm DM}=  f(\phi){\mathcal L}_{\rm DM*} 
\ee
is the nonminimally coupled DM Lagrangian (${\mathcal L}_{\rm DM*}$ denoting the minimally coupled DM Lagrangian). The components of the DM and DE energy-momentum tensors are given by
\bq
T^{\mu\nu}_{\rm DM}&=&\frac{2}{{\sqrt {-g}}} \frac{\delta({\sqrt {-g}}{\mathcal L}_{\rm DM})}{\delta g_{\mu \nu}} = f(\phi) T^{\mu\nu}_{\rm DM*} \,,\label{TDM}\\
T^{\mu\nu}_{\rm DE}&=&\frac{2}{{\sqrt {-g}}} \frac{\delta({\sqrt {-g}}{\mathcal L}_{\rm DE})}{\delta g_{\mu \nu}}\nonumber\\
&=&\nabla^\mu \phi \nabla^\nu \phi +g^{\mu\nu}{\mathcal L}_{\rm DE}\,,
\eq
where $g=\det (g_{\mu\nu})$, $g_{\mu\nu}$ are the components of the metric tensor, and
\be
T^{\mu\nu}_{\rm DM*}=\frac{2}{{\sqrt {-g}}} \frac{\delta({\sqrt {-g}}{\mathcal L}_{\rm DM*})}{\delta g_{\mu \nu}}\,.
\ee
As a consequence of the nonminimal coupling to the quintessence scalar field, the mass $m$ of a DM particle or a compact DM object is a function of $\phi$ with
\be
m(\phi) = f(\phi) m_*\,,
\ee
where $m_*$ is the mass that the particle would have if $f(\phi)=1$. 

\subsection{Scalar field dynamics in a flat Friedmann-Lema\^itre-Robertson-Walker universe}

Consider a flat homogeneous and isotropic universe, described by the Friedmann-Lema\^itre-Robertson-Walker metric. The corresponding line element is given by
\bq
ds^2&=&-dt^2  +a^2[t] d \vec q \cdot d \vec q \nonumber\\
&=& -dt^2  +a^2[t] \left(dq^2 +q^2(d \theta^2 + \sin^2 \theta d \phi^2 )\right)\,,
\eq
where $a[t]$ is the scale factor, $t$ is the physical time, $\vec q$ are comoving Cartesian coordinates, and 
$(q,\theta,\phi)$ are comoving spherical coordinates. If the DM particles are nonrelativistic, then the dynamics of the nonminimally coupled quintessence   scalar field $\phi[t,\vec q\,]$ is given by \cite{Avelino:2015fka}
\be
\Box \phi = -{\ddot \phi} -3 H {\dot \phi} +\nabla^2 \phi  = \alpha - \beta  \sum_i m_i \delta^3[\vec r - {\vec r}_i]\,,
\ee
where $\Box \equiv \nabla_\mu \nabla^\mu$ is the  d'Alembertian, a dot represents a derivative with respect to the physical time $t$, $\vec r = a \vec q$, $m_i$ and $\vec r_i=a \vec q_i$ are, respectively, the masses and positions of the DM particles, $\nabla^2 \phi \equiv  \nabla_{\vec q}^2 \, \phi/a^2$, $\delta^3[\vec r\,]$ is the three-dimensional Dirac delta function, and
\bq
\alpha[\phi] &\equiv& \frac{d V} {d \phi}\,,\\
\beta[\phi] &\equiv& -\frac{d \ln m} {d \phi} = - \frac{d \ln f} {d \phi} \,.
\eq

\section{Quasistatic approximation\label{sec3}}

Consider a single compact DM object of mass $m$ comoving with the expansion of the Universe at ${\vec r}={\vec 0}$. In the quasistatic approximation the field $\phi$ may be written as $\phi[t,{\vec r}\,]= {\bar \phi}[t] + \delta \phi$, with
\bq
{\ddot {\bar \phi}} +3 H {\dot {\bar \phi}} &=& -\alpha [\bar \phi] \label{barphi}\,,\\ 
\nabla^2 \delta \phi &=&  -\beta [\bar \phi] m  [\bar \phi]\delta^3 [\vec r\,]  \label{deltaphi} \,,
\eq
where Eq. (\ref{barphi}) implies that $\bar \phi$ is a function only of the physical time ($\bar \phi=\bar \phi[t]$).
This is expected to be a good approximation as long as
\bq
|\dot {\delta \phi}| &\lsim& |{\dot {\bar  \phi}}|\,, \label{cond1}\\ 
|\delta \phi| &\lsim& \left|\frac{d\ln \alpha}{d\phi}\right|^{-1}\,, \label{cond2}\\
|\delta \phi| &\lsim&  \left|\frac{d\ln (\beta m)}{d\phi}\right|^{-1} \label{cond3}\,.
\eq

The nonzero components of the energy-momentum tensor of the DE field are given approximately by
\bq
{T^t}_t&=& {T^t}_t + \delta {T^t}_t \,,\\
{T^t}_q&=& {T^t}_q+ \delta {T^t}_q \,,\\
{T^q}_q&=& {T^q}_q + \delta {T^q}_q \,,\\
{T^\theta}_\theta&=& {T^\theta}_\theta+ \delta {T^\theta}_\theta \,,\\
{T^\phi}_\phi&=&{T^\theta}_\theta\,,
\eq
with
\bq
{T^t}_t&=&  - \frac12 \dot { \bar \phi}^2  - V[\bar \phi] \,,\\
{T^t}_q &=& 0\,,\\
 {T^q}_q &=&   {T^\theta}_\theta[\bar \phi]= {T^\phi}_\phi[\bar \phi]=\frac12 \dot {\bar  \phi}^2  - V[\bar \phi]\,,
\eq
and
\bq
 \delta {T^t}_t &=& -\frac12 (\delta \phi)'^2 - \alpha[\bar \phi] \delta \phi \,,\\
\delta {T^t}_q &=&  - a\dot {\bar  \phi} \, (\delta \phi)' \,,\\
\delta {T^q}_q &=&  \frac12 (\delta \phi)'^2  - \alpha[\bar \phi] \delta \phi\,,\\
\delta {T^\theta}_\theta &=& \delta {T^\phi}_\phi =  -\frac12 (\delta \phi)'^2  - \alpha[\bar \phi] \delta \phi \,,
\eq
where $ (\delta \phi)' \equiv a^{-1} \partial\phi/\partial q$.

Let us also compute the proper density and pressure associated with the background evolution of the scalar field $\bar  \phi[t]$,
\bq
{\bar \rho}_{\rm DE} &=& - {T^t}_t =\frac12\dot {\bar \phi}^2 + V[\bar \phi]\,,\\
{\bar p}_{\rm DE} &=& \frac13 \left( {T^q}_q+{T^\theta}_\theta+{T^\phi}_\phi\right)=\frac12\dot {\bar \phi}^2 - V[\bar \phi]\,,
\eq
and define the current values of the corresponding equation of state and fractional energy density parameters:
\bq
\bar w&\equiv&{\bar p}_{\rm DE}/{\bar \rho}_{\rm DE}\,,\\
{\bar \Omega}_{\rm DE}&\equiv&\frac{8 \pi G {\bar \rho}_{\rm DE}}{3 H^2}\,.
\eq
Notice that
\bq
{\dot {\bar \phi}}^{\,2} &=& (1+\bar w) {\bar \rho}_{\rm DE} = \frac{3 H^2 (1+\bar w) {\bar \Omega}_{\rm DE}}{8 \pi G}  \label{barphi2}\,,\\
V(\phi) &=& (1-\bar w) {\bar \rho}_{\rm DE} = \frac{3 H^2 (1-\bar w) {\bar \Omega}_{\rm DE}}{8 \pi G} \,. \label{barphi3}
\eq

\subsection{Quasistatic solution}

The quasistatic solution to Eq. (\ref{deltaphi}) outside a compact DM object of radius $R$ is given by 
\be
\delta \phi [t,r]=\delta \phi [t,R]\frac{R}{r} \,, \qquad r=|\vec r\,| \ge R\,,\label{quasistatic}
\ee
where
\be
\delta \phi [t,R]=\frac{\beta m }{4\pi   R}\,.\label{quasistatic1}
\ee
The DE flux towards the interior of a sphere of radius $r \ge R$ centered on  the DM compact object is
\bq
- 4 \pi r^2 {\dot {\bar \phi}}\, (\delta \phi)' = \beta \, {\dot {\bar \phi}} \, m = {\dot m} \,,
\eq
thus accounting for the change of the object's mass.

Calculating the partial derivative of $\delta \phi$ with respect to the physical time $t$ (at fixed $q=r/a[t]$) one obtains
\be
\dot {\delta \phi}=\left(\frac{d\ln (\beta m)}{d\phi}\dot  {\bar \phi}-H\right) \delta \phi\,,
\ee
where $H \equiv \dot a /a$ is the Hubble parameter.
The conditions given in Eqs. (\ref{cond1}) and (\ref{cond3}) thus imply that the quasistatic approximation is expected to be a good approximation for
\be
(\dot {\delta \phi})^2 \sim H^2 \delta \phi^2  \lsim  {\dot {\bar \phi}}^{\,2} \,,
\ee
or, equivalently,
\be
\frac{r}{R} \gsim 2 \frac{\boldsymbol{\beta}^2}{|k|} \varphi \,.
\label{approxcond}
\ee
Here, $\boldsymbol{\beta} $ is the nonminimal coupling strength parameter defined by
\be
\boldsymbol{\beta}\equiv \frac{\beta}{\sqrt {8\pi G}} \,, \label{betaadim}
\ee
\be
k[a]\equiv\frac{d \ln m} {d \ln a}=-\frac{\beta \dot {\bar \phi}}{H} =- \boldsymbol{\beta} \sqrt{3(1+\bar w) {\bar \Omega}_{\rm DE}}\label{ka}
\ee
is a related cosmological coupling strength parameter also used often in the literature, and 
\be
\varphi\equiv \frac{Gm}{R}
\ee
is the Newtonian gravitational potential at the surface of the compact DM object. Equation (\ref{approxcond}) implies that the condition
\be
|k| \gsim 2 \boldsymbol{\beta}^2 \varphi
\ee
is required in order to ensure that the quasistatic approximation is valid for all $r > R$. 

\section{DE halos\label{sec4}}

The energy density fluctuations of the quintessence scalar field $\delta \rho =- \delta {T^t}_t= (\delta \phi)'^2/2 + \alpha[\bar \phi] \delta \phi$ are the sum of two contributions: (1) the one directly associated with the scalar field gradients $(\delta \phi)'^2/2 \propto r^{-4}$ and (2) the one associated with spatial variations of the potential energy $\alpha[\bar \phi] \delta \phi \propto r^{-1}$. The first one provides a contribution that is essentially localized in and around the particle, while the second one is strongly dependent on how one defines the halo radius $r_{\rm halo}$. In fact, the energy perturbation inside a sphere of radius $r_{\rm halo}$ centered on the compact DM object associated with contribution 2 is given by
\bq
\delta E_V [r_{\rm halo}] &=& 4 \pi \int_0^{r_{\rm halo}} \alpha [\bar \phi] \delta \phi r^2 dr  \nonumber \\
&=&2 \pi \alpha [\bar \phi] \delta \phi [t,R] R r_{\rm halo}^2  \nonumber \\
&=& \frac32\left( \frac{\alpha [\bar \phi] \delta \phi[t,R]}{V[\bar \phi]} \right) \frac{R}{r_{\rm halo}} E_{V},,
\eq
where $E_V = 4\pi  r_{\rm halo}^3 V [\bar \phi]/3$. Hence, if $|\alpha[\bar \phi]  \delta \phi [t,R]| \ll V[\bar \phi]$ or $R \ll r_{\rm halo}$, then this contribution may be neglected. Therefore, in this paper we shall focus only on the contribution directly associated with the scalar field gradients. 

\subsection{Halo energy}

The total energy associated with the scalar field gradients outside a sphere of radius $R$ centered at $r=0$ is given by
\be
E^{\rm out}_{\rm G}[t] = 2 \pi \int_R^\infty  (\delta \phi)'^2  r^2 dr =\boldsymbol{\beta}^2 \varphi \, m \,, \label{EG}
\ee
which will be our best estimate of the total energy of a DE halo ($E_{\rm halo} \sim E^{\rm out}_{\rm G}$).
For $\varphi \equiv Gm/R \gsim 1$ this result should be taken as a rough estimate, since the impact of the local gravitational field on the dynamics of the quintessence scalar field has not been taken into account. 

\subsection{Halo radius}

The radius of a DE halo may be determined by defining the halo as the region where the energy density associated with the scalar field gradients exceeds the background energy density or, equivalently, $(\delta \phi)'^2/2 >  3H^2/(8\pi G)$. This happens for values of $r$ smaller than
\be
r_{\rm halo}[t]= \left(\sqrt \frac23  \boldsymbol{\beta} \,\varphi \frac{R}{H}\right)^{1/2}\,.
\ee
Here we implicitly assume that $R <  \sqrt {2/3} \, \boldsymbol{\beta} \varphi H^{-1}$ or, equivalently, that $r_{\rm halo} > R$.

\subsection{Equation of state}

Within a DE halo the DE equation of state parameter,
\be
w_{\rm DE}\equiv   \frac{p_{\rm DE}}{\rho_{\rm DE}} =  \frac{{\bar p}_{\rm DE}+{\delta p}_{\rm DE}} {{\bar \rho}_{\rm DE}+{\delta \rho}_{\rm DE}} \,,
\ee
is dominated by the contribution of the scalar field gradients to the DE proper density and pressure. These satisfy  $\delta p_{\rm G}=-(\delta \phi)'^2/6=-\delta \rho_{\rm G}/3$ or, equivalently,
\be
w_{\rm G}\equiv  \frac{\delta p_{\rm G}}{\delta \rho_{\rm G}} = -\frac13\,.
\ee
Hence, the DE equation of state parameter is expected to be close to $-1/3$ within an halo, especially in its central regions.

\subsection{DE perturbations inside compact DM objects}

The knowledge of the density profile inside a compact DM object would be required in order to accurately compute the energy $E^{\rm in}_{\rm G}$ associated with the scalar field gradients for $r < R$ --- notice that this contribution would vanish if and only if all the mass was located at the surface of the object, which would be utterly unrealistic. Here we will provide a rough estimate of $E^{\rm in}_{\rm G}$, assuming that $Gm/r$ is always significantly smaller than unity and that the DM energy density distribution for $r < R$ is uniform. In this case, the quasistatic solution to Eq. (\ref{deltaphi}) is given by 
\be
\delta \phi [t,r]=\frac32\delta \phi [t,R]\left(1-\frac13 \left(\frac{r}{R} \right)^2\right)\,.\label{quasistatic1}
\ee
The total energy directly associated to the scalar field gradients inside the compact DM object would then be equal to
\be
E^{\rm in}_{\rm G}[t]= 2 \pi \int_0^R  (\delta \phi)'^2  r^2 dr = \frac15 E^{\rm out}_{\rm G}[t]\,,
\ee
so that $\sim 17 \%$ of the energy associated with the scalar field gradients would be located inside the DM object. In general one would expect the DM energy density to be a decreasing function of $r$. Therefore, this should be regarded as a lower limit on $E^{\rm in}_{\rm G}$.

\section{Discussion and Conclusions}\label{sec:conc}

In this paper we characterized the properties of DE concentrations that are expected to form in and around compact DM objects as a consequence of a coupling of the mass of DM particles to a standard quintessence scalar field. We determined the dependence of the energy and radius of a DE halo on the nonminimal coupling strength and Hubble parameters, and on the mass and radius of the associated compact DM object. We have also shown that deep inside a DE halo the equation of state parameter is close to -1/3 and estimated the contribution of the DE energy perturbations inside DM objects. 

Although the results presented in this paper were obtained considering a single comoving compact DM object, they should also hold in the case of a network of nonrelativistic DM objects. However, in this case the background evolution of the DE scalar field will be affected by the energy transfer between DE to DM. Although this effect needs to be considered for an accurate characterization of the background dynamics of the DE scalar field, it is not expected to significantly change our main results. Also, our results were obtained in the context of the quasistatic approximation, assuming that local contributions to the evolution of the mass of dark matter particles can be neglected. If that is not the case a breakdown of the quasistatic approximation is expected, which can be associated with a significant transfer of linear-momentum between moving DM particles and the DE scalar field \cite{Avelino:2015fka}.

Scalar field gradients associated with the coupling of the mass of DM particles to a DE scalar field give rise to attractive fifth forces between DM particles whose strength is equal to $2 \boldsymbol{\beta}^2$ times that of gravitational forces. In addition to these, energy-momentum conservation in general relativity requires that any change of the proper mass of a compact DM object associated with the nonminimal coupling to the DE scalar field should be compensated by a corresponding decrease of its speed with respect to the local cosmological frame,  assuming that backreaction effects associated with a significant transfer of linear momentum from moving DM particles to the DE scalar field can be neglected (which is expected to be the case as long as the quasistatic approximation holds everywhere). This effect essentially changes the strength of the velocity dependent cosmological damping of the speed of compact DM objects by a factor of $1+k$. 

The velocity dependent forces and fifth forces in coupled DE energy models have an impact on the growth of cosmic structures which can  be constrained observationally. A tomographic analysis of coupled DE has been performed in \cite{Goh:2022gxo}, where a significant redshift dependence of the constraints on the coupling strength has been found (assuming, for simplicity, a non-negative coupling strength parameter): $\boldsymbol{\beta} \lsim 0.1$ at $z<5$, and $\boldsymbol{\beta}\lsim 0.05$ in the redshift range $5<z<500$, while  $\boldsymbol{\beta}\lsim 0.02$ for a constant coupling case (at $68\%$ confidence level). This imposes stringent constraints on the ratio between the energy of a DE halo and the mass of the associated compact DM object: $E_{\rm halo}/m \lsim 0.01  \varphi$ at $z<5$. It also precludes any substantial contribution (i.e.,  exceeding a subpercent level) of DE halos to the DE equation of state parameter. Whether these limits can be relaxed by considering broader families of coupled DE models will be the subject of future work.

\begin{acknowledgments}
	
We thank Lara Sousa, Rui Azevedo, Vasco Ferreira, and David Gr{\"u}ber for many enlightening discussions. We acknowledge the support by Fundação para a Ciência e a Tecnologia (FCT) through the research Grants No. UIDB/04434/2020 and No. UIDP/04434/2020. This work was also supported by FCT through the R$\&$D project 2022.03495.PTDC - Uncovering the nature of cosmic strings.

\end{acknowledgments}
 
\bibliography{DEHalos}
 	
 \end{document}